\documentclass{biophys_letter}
\usepackage{dcolumn,color,helvet,times}
\usepackage{amsmath,lastpage,bm,textcomp}
\usepackage{multicol}
\jno{kxl014} 
\gridframe{N}
\cropmark{N}
\numberwithin{equation}{section}
        
\doi{doi: 10.1529/biophysj.xxxx}

\newcommand{\CLH}[1]{{\bf CLH: #1}}
\newcommand{\LATER}[1]{{}}
\newcommand{\CLHdone}[1]{{}}

\newcommand{\ttauL}{{\tilde{\tau}_L}}
\newcommand{\phaseph}{{\scriptstyle \rm ph}}
\newcommand{\phaseI}{{\scriptstyle \rm I}}
\newcommand{\phaseII}{{\scriptstyle \rm II}}
\newcommand{\phaseIII}{{\scriptstyle \rm III}}
\newcommand{\phaseIV}{{\scriptstyle \rm IV}}

\begin{document}


\setcounter{page}{1} 
\title{Propagating left/right asymmetry in the zebrafish embryo:
one-dimensional model}

\author{Hanrong Chen,* C.~L.~Henley*, and B. Xu\authdagger
\LATER{R. D. Burdine\authdagger}}
\address{*Dept. of Physics, Cornell University, Ithaca NY 14853 USA and 
{\addrdagger} Dept. of Molecular Biology, Princeton University,
Princeton NJ 0854x, USA}

\maketitle

\pagestyle{headings}

\markboth{Biophysical Journal: Biophysical Letters}{Biophysical Journal: Biophysical Letters} 


\begin{abstract}
{During embryonic development in vertebrates, left-right (L/R) asymmetry is reliably generated by a conserved mechanism: 
a L/R asymmetric signal is transmitted from the embryonic node to other parts of the embryo 
by the L/R asymmetric expression and diffusion of the TGF-$\beta$ related proteins Nodal and Lefty
via propagating gene expression fronts in the lateral plate mesoderm (LPM) and midline. 
In zebrafish embryos, Nodal and Lefty expression can only occur along 3 narrow stripes that express the co-receptor 
\emph{one-eyed pinhead} (oep): Nodal along stripes in the left and right LPM, and Lefty along the midline.
In wild-type embryos, Nodal is only expressed in the left LPM but not the right, because of 
inhibition by Lefty from the midline; however, bilateral Nodal expression occurs in loss-of-handedness mutants.
A two-dimensional model of the zebrafish embryo predicts this loss of L/R asymmetry in oep mutants \cite{henley-xu-burdine}. 
In this paper, we simplify this two-dimensional picture to 
a one-dimensional model of Nodal and Lefty front propagation along the oep-expressing stripes.
We represent Nodal and Lefty production by step functions that turn on when a linear function 
of Nodal and Lefty densities crosses a threshold.
We do a parameter exploration of front propagation behavior, and find the existence of \emph{pinned} intervals, 
along which the linear function underlying production is pinned to the threshold.
Finally, we find parameter regimes for which spatially uniform oscillating solutions are possible.
}
{Received for publication xx yy 2011 and in final form xx yy 2012.}
{Address reprint requests and inquiries to Christopher L. Henley, E-mail: 
clh@ccmr.cornell.edu}
\end{abstract}

\vspace*{2.7pt}
\begin{multicols}{2}


\section{Introduction}

In vertebrates, organs such as the heart and brain develop in an invariant left-right
(L/R) asymmetric fashion, with \emph{situs inversus}, or complete
organ mirror-reversal, being a rare occurrence \cite{hamada,levin,raya}.
The best-understood mechanism by which L/R bias is first
established is that special cilia around the embyronic node 
drive a leftwards fluid flow \cite{mcgrath}. Disrupting, or reversing this flow results 
in randomization \cite{nonaka}, or reversal \cite{nonaka-hamada} of organ asymmetry.
The cilia appear to be conserved in vertebrates \cite{essner-tabin-yost}, 
but it is controversial whether the nodal flow mechanism 
is decisive in all vertebrates, since asymmetry of some other origin 
(involving electrical potentials due to polarized gap junctions) 
is observable before the cilia even begin moving, and this 
may determine organ asymmetry in {\it Xenopus} \cite{adams-levin}.

Our concern in this paper is with the subsequent propagation of this L/R signal from the node
to the rest of the embryo. It is first transmitted to the lateral plate mesoderm
(LPM), resulting in L/R asymmetric expression of the TGF-$\beta$
related signaling molecules Nodal and Lefty \cite{kawasumi-hamada,brennan}.
Nodal is expressed exclusively in the left LPM, and Lefty in the midline, along
gene expression fronts that propagate from the posterior to the anterior of the embryo
from approximately the 12 to 20 somite stages  \cite{wang-yost}.
This propagation depends on the diffusion of Nodal and Lefty, as well as their
production rates which are promoted or inhibited by these same molecules. 
Subsequently, downstream genes (such as Pitx2) are activated and direct
L/R asymmetric organogenesis \cite{ryan-nature}. 

Nodal and Lefty constitute a classic activator-inhibitor system that is highly conserved among vertebrates \cite{hamada}. 
Such a model was introduced and studied for the 
initiation of Nodal and Lefty expression in the LPM and midline, 
respectively, at the level of the node in the mouse \cite{nakamura-hamada}.
This was spatialy one-dimensional, representing the L/R axis, and
did not attempt to model the longitudinal propagation of the 
L/R signal. 

In this paper, we extend the ideas in this model to the
zebrafish embryo, in which Nodal and Lefty gene expression 
only occurs in cells carrying the co-receptor
 \emph{one-eyed pinhead} (oep) found on narrow stripes 3-5 cells wide in the midline
 and LPM \cite{xu}. (Note that the Nodal analogue in the zebrafish is called ``Southpaw'', 
 but we will call it ``Nodal'' for consistency with the terminology 
of Ref.~\cite{nakamura-hamada}). 
An advantage of this system is that it is
quasi-one-dimensional in the longitudinal direction, because the
most important variables are Nodal and Lefty production
on these 3 stripes, and the propagation of their gene expression fronts is clearly 
unidirectional in the posterior-to-anterior direction.

\doiline
\end{multicols}
\twocolumn


A two-dimensional model of the Nodal/Lefty system in 
the zebrafish embryo has been set up and 
simulated~\cite{xu,henley-xu-burdine}, as we review
in Section~\ref{sec:2D-model}. In Sections~\ref{sec:1D-model} and~\ref{sec:results}, 
we work through the consequences of a one-dimensional
idealization of it, representing not the L/R axis  (as in \cite{nakamura-hamada}) but the anterior-posterior axis.
We idealize Nodal and Lefty production to have step-function turn-ons, when a threshold function
linear in their concentrations crosses a threshold.
We classify all possible behaviors in the limit that Nodal is unaffected by Lefty, and 
demonstrate the existence of intervals along which the Lefty profile must be \emph{pinned} to that of Nodal.
In Section~\ref{sec:oscillating}, we find parameter regimes that allow uniform spatial oscillations to occur.

\section{Two-dimensional model}
\label{sec:2D-model}

In this section, we outline what would be a minimal model
for the two-dimensional Nodal/Lefty system.  This will
motivate the simplification we make, in 
section~\ref{sec:1D-model}, to a one dimensional 
model of similar form.  
The degrees of freedom in the model we use here are
only the concentrations $N(r,t)$ of Nodal and $L(r,t)$ of
Lefty, as found unbound in the extracellular fluid.
In this section we let $r$ represent a two-component
position $(x,y)$ on the embryo, which
is practically a flat two-dimensional layer enveloping the yolk cell; in later
sections on the one-dimensional model it will be replaced by the variable $y$.
Also, $t$ is the time.

\begin{figure}[t!]\vspace*{3pt}
 \centering{\includegraphics[width=3.3in]{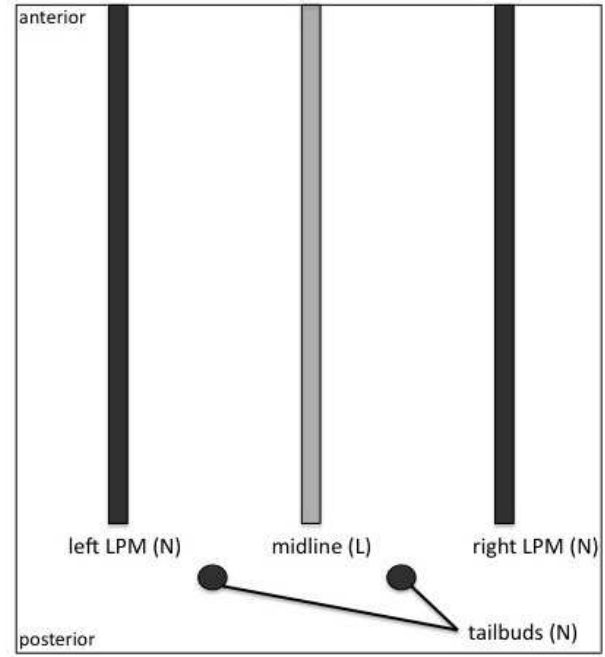}}\vspace*{-5.5pt}
\caption{
\label{fig:layout}
Layout of the 2D problem \cite{henley-xu-burdine}. Nodal is only expressed in the dark shaded regions, and Lefty
is only expressed in the light shaded region. Initially the left tailbud expresses more Nodal than the right.
}\vspace*{-8pt}
\end{figure}

The geometry is idealized so that the three oep-expressing stripes 
extend indefinitely for $y>0$, and are straight, parallel, and
of unvarying widths, as shown in Fig.~\ref{fig:layout}.  That means that, away from the baseline
at $y=0$, the model system has a translational invariance.
That permits one to talk in a mathematically precise way about 
a limiting behavior in the model, in particular \emph{uniform 
motion of a front}.  Whether this is pertinent to the
actual embryo depends on whether initial transients persist
for a short time compared to the stages in which the L/R
signal propagates, and for a short distance compared to
the inter-stripe spacing or to the embryo's length, which we know to be the case \cite{wang-yost}.

Lefty is produced at a rate $s_L(r,t)$ [concentration 
per unit time per unit area] which can be nonzero
only at points lying within the midline stripe,  
since only those cells can produce Lefty.
The function $s_L(r,t)$ depends, in our simplified
model, only on the local, instantaneous concentrations 
of Nodal and Lefty; the exact functional dependence 
is specified in Sec.~\ref{sec:N-L-production}.
More precisely, the production function depending
on $N(r,t)$ and $L(r,t)$ should rather be the 
transcription rate of the Lefty mRNA, while
the source rate $s_L(r,t)$ of Lefty protein is in turn proportional 
to the mRNA concentration, and thus should lag the  
production function we use by roughly the mRNA lifetime.
(Indeed, a model including that lag was successfully
fitted to quantitative data in \cite{xu}, Chapter 3.)
But in this paper, we assume that the source rate is
proportional to the instantaneous production function;
that is equivalent to assuming the Lefty mRNA is transcribed
copiously and decays quickly compared to other time scales.

Similarly, Nodal is produced at a rate $s_N(r,t)$ which
is nonzero only within stripes in the left LPM and right LPM, being described
by a production function of the same form as that of Lefty
in the midline, and with the same
production function in the right and left LPM, so \emph{no
prior asymmetry} is assumed.  As with Lefty, we have 
made an approximation of short-lived mRNA so that
the mRNA content is not treated as an independent 
variable.  

In the case of Nodal, there is a second
place where $s_N(r,t)$ is nonzero, which are the two
tailbuds, located as shown in Figure~\ref{fig:layout}.
This production is {\it not} modulated by $N(r,t)$
or by $L(r,t)$, and  may be assumed constant in time
for simplicity (in reality it gradually diminishes \cite{xu}).
The tailbud production has some left-right asymmetry, 
which in our model is responsible for \emph{all} downstream
asymmetries. The origin of the asymmetry is imagined to 
be in the nearby Kupffer's vesicle~\cite{essner-KV} 
(the organ in which the special cilia are found in zebrafish).

\subsection{Diffusion behavior}\LATER{and parameters}

Once produced, Nodal and Lefty are assumed to be passive,
noninteracting densities that satisfy a diffusion equation
with diffusion constants $D_N$ and $D_L$ respectively.
Furthermore, they are assumed to have degradation rates,
respectively $\tau_N^{-1}$ and $\tau_L^{-1}$.  Although
one expects the two signaling molecules to have similar 
values of these parameters, we allow them to be different
in general, because (as we derive in the results sections)
this inequality induces qualitiatively {\it different} behaviors.
(The differential equations incorporating these
parameters are exactly like Eqs.~\eqref{eq:dN/dt} 
and~\eqref{eq:dL/dt} 
of the one-dimensional model,
except the second derivative with respect to $y$ is replaced 
by the two-dimensional divergence.)

Characteristic length and velocity scales can be constructed
from the diffusion constant and degradation times:
 \begin{equation}
        l_{N}=\sqrt{D_{N}\tau_{N}};
 \end{equation}
 \begin{equation}
  v_{N}=\sqrt{\frac{D_{N}}{\tau_{N}}},
 \end{equation}
and we define $l_{L}$ and $v_{L}$ similarly.
These scales determine the typical range in space 
that the signal molecules travel, and the typical
speed of the production fronts we shall study.

\subsection{Conditions for Nodal and Lefty production}
\label{sec:N-L-production}

A key property determining the behavior is that 
expression (and production) of either Nodal or Lefty is 
\emph{promoted} by the presence of Nodal and \emph{inhibited} by
the presence of Lefty.   Thus, both production functions
$s_N(r,t)$ and $s_L(r,t)$ are increasing as a function  of
$N$  and decreasing as a function of $L$.
As we explain next, one can consider several
versions of our model with different functional forms 
for the production functions.  

Let us first lay out how production is actually regulated.
Along the LPM and midline stripes, Nodal and Lefty protein bind to
the plasma membrane with the help of oep co-receptors \cite{xu}; 
we may say ``oep" is in one of three states: 
unbound, Nodal-bound or Lefty-bound. 
We assume that only a negligible fraction of all Nodal and Lefty
produced are thus sequestered by the receptors, so that
uptake or release of the signaling molecules is not included in
$s_N(r,t)$ and $s_L(r,t)$.

In turn, Nodal-bound oep induces a signaling cascade that
promotes transcription of Nodal (in the LPM) or Lefty (on the midline). 
The fraction of Nodal-bound oep, $\phi(N,L)$, is
 \begin{equation}
 \phi(N,L)=\frac{N}{k^{-1}+N+L}
 \label{eq:phi}
 \end{equation}
The oep receptors on the midline are believed to be identical to
those on the LPM stripes, 
so we use the {\it same} function $\phi(N,L)$ in both stripes 
with the same equilibrium constant $k$ (where $k^{-1}$
has the units of concentration).

The Nodal production function $s_N$ is assumed to be 
proportional to a Hill function
\begin{equation}
  f_N(\phi)=\frac{1}{1+ (\phi_{*N}/\phi)^{n_H}}.
\label{eq:Hill}
\end{equation}
This is a rounded step from 0 to 1 centered at a threshold parameter
$\phi_{*N}$; the step gets sharper
as the Hill exponent $n_H$ grows large.
A microscopic rationalization for the form in Eq.~\eqref{eq:Hill}
would be that the Nodal gene's regulatory region cooperatively
binds $n_H$ copies of the signal or transcription factor 
induced by Nodal-bound oep.
(Ref.~\cite{nakamura-hamada} used a different production function
than \eqref{eq:Hill} that also approximates a step function with a parameter
controlling the sharpness.)

The actual production rate is 
\begin{equation}
s_N(r,t)=s_{N0} f_N(\phi(N,L))
\end{equation}
where $s_{N0}$ is the maximum, or saturated, production rate.
A similar function $f_L(\phi)$, with 
(it is expected) a rather different threshold
$\phi_{*L}$, and a saturated production rate $s_{L0}$,
controls Lefty production.

Part of the motivation for including the intermediate function
$\phi$ in the model is to represent mutants in which oep is
under- or over-expressed.  That would have the effect
of multiplying $\phi$ in \eqref{eq:Hill} by a constant, or
equivalently of dividing $\phi_{*N}$ and $\phi_{*L}$ by
that constant~\cite{xu}.

We adopt two simplifying assumptions, either of which could
be made independently of the other.
The first pertains to the threshold for turning production on or off.
In place of the nonlinear $\phi$ function \eqref{eq:phi},
we follow Ref.~\cite{nakamura-hamada} and 
instead use two threshold functions
linear in the concentrations:
\begin{subequations}
\label{eq:Cij}
\begin{equation}
C_{N}(N,L)=C_{NN}N-C_{NL}L
\label{eq:C_N}
\end{equation}
\begin{equation}
C_{L}(N,L)=C_{LN}N-C_{LL}L
\label{eq:C_L}
\end{equation}
\end{subequations}
where $N=N(r,t)$ and $L=L(r,t)$
are Nodal and Lefty concentration, respectively,
and $C_{NN}$, $C_{NL}$, $C_{LN}$, and $C_{LL}$ are constants.
Production of $N$ or $L$ respectively turns on when 
this function exceeds a threshold parameter
$C_{N*}$ or $C_{L*}$, which should be positive,
since a certain concentration of Nodal is needed to turn on
either production, even in the absence of Lefty.
We set $C_{N*}\equiv C_{L*}\equiv  1$
without loss of generality by scaling the four coefficients
in Eqs.~\eqref{eq:Cij}.

The relation of Eqs.~\eqref{eq:Cij} to the $\phi$ function 
in \eqref{eq:phi} is that the condition to
be over threshold, $\phi>\phi_{*N}$ or $\phi>\phi_{*L}$,
 is (by inserting Eq.~\eqref{eq:phi})
mathematically equivalent to 
\begin{subequations}
\label{eq:phi-to-linear}
\begin{equation}
   k (\phi_{*N}^{-1}-1) N - kL  > 1;
\end{equation}
\begin{equation}
   k (\phi_{*L}^{-1}-1) N - kL  > 1.
\end{equation}
\end{subequations}
Clearly, we can (roughly) identify the coefficients in 
\eqref{eq:phi-to-linear} with those in \eqref{eq:Cij},
in particular this viewpoint implies $C_{NL}\equiv C_{LL}$.

Note that although the condition to be at threshold is captured
exactly by the linear functions, away from threshold the production 
will depend on $N$ and $L$ in a different way when $n_H<\infty$.
The linear form of Eqs.~\eqref{eq:C_N} and~\eqref{eq:C_L} 
will be more convenient mathematically.

The second simplification we make is to assume 
$n_H\rightarrow\infty$ in \eqref{eq:Hill},
i.e. a step-function turn-on of the production. 
In this case,
Nodal production $s_{N}(r,t)$ and Lefty production
$s_{L}(r,t)$ are given by:

\begin{subequations}
\label{eq:prod_N_L}
\begin{equation}
\begin{cases}
s_{N}(r,t)=0 & ,\: C_{N}<1\\
0\leq s_{N}(r,t)\leq s_{N0} & ,\: C_{N}=1\\
s_{N}(r,t)=s_{N0} & ,\: C_{N}> 1;
\end{cases}\label{eq:prod_N}
\end{equation}
and
\begin{equation}
\begin{cases}
s_{L}(r,t)=0 & ,\: C_{L}<1\\
0\leq s_{L}(r,t)\leq s_{L0} & ,\: C_{L}=1\\
s_{L}(r,t)=s_{L0} & ,\: C_{L}>1.
\end{cases}\label{eq:prod_L}
\end{equation}
\end{subequations}

\subsection{Behavior of the two-dimensional model and comparison with experiments}

The model initializes with zero Nodal or Lefty concentration in the domain (Fig.~\ref{fig:layout}), 
and with the 2 tailbuds producing Nodal that diffuses to the base of the 3 stripes. 
The left tailbud produces more than the right, which is responsible for all downstream asymmetries.
Successful initiation at the base of the stripes requires that the Nodal produced in the tailbuds
 is sufficient to start self-sustaining Nodal production in the left LPM, but not the right LPM;
 production begins when thresholds given in \eqref{eq:prod_N_L} are exceeded. 
 The times at which this occurs differs for all 3 stripes depending on the difference between
$\phi_{*N}$ and $\phi_{*L}$, the amounts of Nodal produced by the left and right tailbuds, 
as well as the position of the tailbuds (closer to or farther from
the midline, in a L/R symmetric way). Lefty, once produced on the midline, 
diffuses to the left and right LPM in equal amounts.
Thus, the threshold function \eqref{eq:C_N} is constrained such that 
Nodal production turns off in the right LPM due to Lefty inhibition,
but stays on in the left because of greater $N$ there. 
This occurs in simulations for widely-varying parameter sets, 
for differences in tailbud production rates of 30\%. 
As this asymmetry is decreased, the coefficients in \eqref{eq:C_N}, 
as well as the geometry of the problem,
must be tuned more finely for successful initiation to occur. 
\LATER{has the difference in Nodal production between the left and right tailbuds been measured?}

Experiments in zebrafish find that after an initial phase the wavefronts
of Nodal and Lefty gene expression proceed at a fixed speed \cite{xu,wang-yost},
and indeed this is reproduced in simulations:
once initiated in a self-sustaining way, Nodal production on the left LPM switches on from the base upwards, 
and the leading edge of production moves at an asymptotically fixed rate along the stripe, followed by 
a front of Lefty production, moving at the same asymptotic rate on the midline.

The main role of Nodal and Lefty in this model is to transmit the L/R asymmetric signal from the tailbud.
Lefty in particular acts as a "barrier" preventing initiation of Nodal production in the right stripe: 
the Lefty concentration there is sufficiently high
that Nodal diffusing from the left LPM is insufficient to initiate production there.
Another necessary condition for successful propagation is that 
the function \eqref{eq:C_L} must not fall below 1 along the midline,
or else Lefty production will cease, and Nodal produced in the left LPM will not be 
prevented from diffusing to the right LPM and
initiating production further up the right stripe, leading to loss of L/R asymmetric propagation.
Indeed, in Lefty knockout mutants, bilateral Nodal expression is observed \cite{wang-yost,nakamura-hamada}.


We intended to model the space and time-dependent behaviors
in this model to, ultimately,
\begin{itemize}
\item[(i)]
confirm the correctness of the basic picture by
its qualitative agreement with experiments;
\item[(ii)]
explain, and predict, the phenotypes of various mutations
involving this system;
\item[(iii)] 
characterize the robustness of how the initial
bias in the tailbud is amplified (as is manifested in the
error rate);
\item[(iv)]
show how the numerical values of the various parameters
can be inferred (or as least bounded) on the basis of
experiments.
\end{itemize}
However, while it is understood in principle how we get a
moving front, the mathematical formulation
involves convolutions in two dimensions and 
does not include closed forms of simple functions.
Furthermore, simulations suggested various possible
regimes in which production might turn off again
after the initial passage of a pulse, or in which
the midline Lefty production had a gradual onset
in space, much less sharp than the Nodal front.
In order to analytically represent the functional form,
and to comprehensively classify all regimes of
the asymptotic behavior, we turned to a one dimensional
model, which is the main focus of the rest of this paper.

\section{One-dimensional model}
\label{sec:1D-model}

The 2D model may be simplified to 1D by making the following simplifications.
Since Nodal and Lefty expression occur on 
narrow, parallel stripes, we may replace the finite width of each stripe by a 1D line.
Also, the concentration profile on one stripe is transmitted (with a diffusive lag) to another
through the intervening space; this space can be replaced by lines as well, and we end up with a 
5-line 1D model with diffusion between stripes given by
\begin{equation}
N_i(y,t)=k_D N_{i-1}(y,t-\Delta t)+k_D N_{i+1}(y,t-\Delta t)+...
\end{equation}
where $N_i$ and $N_{i\pm1}$ are Nodal concentrations on adjacent stripes, and $k_D$ and $\Delta t$ describe
the rate of diffusion.

In this paper, we will only focus on the front propagation of Nodal and Lefty gene expression
in the steady state, in which case we further neglect inter-stripe diffusion and thus simplify the system to 1D.

\subsection{Diffusion equations}

The partial differential equations for Nodal and Lefty are:

\begin{subequations}
\label{eq:d/dt}
\begin{equation}
\frac{\partial N(y,t)}{\partial t}=D_{N}\frac{\partial^{2}N(y,t)}{\partial y^{2}}-\frac{1}{\tau_{N}}N(y,t)+s_{N}(y,t)
\label{eq:dN/dt}
\end{equation}
\begin{equation}
\frac{\partial L(y,t)}{\partial t}=D_{L}\frac{\partial^{2}L(y,t)}{\partial y^{2}}-\frac{1}{\tau_{L}}L(y,t)+s_{L}(y,t)
\label{eq:dL/dt}
\end{equation}
\end{subequations}

The equations appear linear at first glance, but the 
$s_N$ and $s_L$ terms are in fact nonlinear, since 
(according to \eqref{eq:prod_N} and \eqref{eq:prod_L})
they depend on $(N,L)$ via step functions.
Nevertheless,
if either $C_N(y)<1$ or $C_N(y)>1$ throughout an interval,
then $s_N(y)$ is constant (either 0 or $s_{N0}$) in that
interval, in which case the differential equation for $N(y)$
may be solved, and similarly for $L(y)$.
Evidently, the key parameter in such a solution
is the position $y$ at which the threshold is crossed
and production turns on or off.  

Surprisingly, it is
also generically possible that the threshold function
is pinned at 1 throughout a finite interval (see Sec.~\ref{sec:pinned}). Within such an
interval, the production function is indefinite (the
middle case in Eqs.~\eqref{eq:prod_N} and \eqref{eq:prod_L})
and the diffusion equation can no longer be used to find
the solution.  However, the threshold
condition becomes an equality interval and gives 
the solution in such intervals.  The complications
in our results have to do mostly with handling such
``pinned'' intervals.


The simplest possible solution is for constant, homogeneous production,
e.g. $s_N(y,t)=s_{N0}$ for all $(y,t)$ for Nodal.
The solution is $N(y,t)=N_{0}$, or 
$L(y,t)=L_{0}$ in the analogous case for Lefty,
where
\begin{subequations}
   \begin{equation}
         N_0 \equiv\tau_{N}s_{N0};
   \end{equation}
   \begin{equation}
         L_0 \equiv\tau_{L}s_{L0}.
   \end{equation}
\end{subequations}
The middle of any interval of production looks
locally like a piece of this homogeneous system, so
it is not surprising that $N_0$ and $L_0$ serve as a reference
level for all solutions.  In particular, since $s_N(y,t)\leq s_{N0}$
everywhere, any solution must have $N(y,t)\leq N_0$ everywhere,
and similarly for $L(y,t)$.

\subsection{The traveling wavefront solution}
\label{sec:traveling-wavefront}

Experiments show the Nodal front is typically ahead of the 
Lefty front~\cite{wang-yost,xu}.  
If it is sufficiently far ahead, the Lefty concentration there 
is negligible, and it is sufficient to consider a Nodal-only
situation, in which the front of Nodal production moves at constant speed $v$:
\begin{equation}
s_{N}(y,t)=\begin{cases}
s_{N0} & ,\: y<vt\\
0 & ,\: y>vt .
\end{cases}\label{eq:wave prod}
\end{equation}
The front shows a constant shape to a viewer
traveling at velocity $v$, i.e. $N(y,t)=N(y-vt)$.
When this is inserted into Eq.~\eqref{eq:dN/dt},
it becomes an ordinary, second-order, linear differential
equation, which is solved by an exponential:
\begin{equation}
N(y-vt)=\begin{cases}
N_{0}[1-g_{N}e^{\kappa_{N}(y-vt)}] & ,\: y-vt<0\\
N_{0}(1-g_{N})e^{-\kappa'_{N}(y-vt)} & ,\: y-vt>0.
\end{cases}
\label{eq:N-for-step-function}
\end{equation}
Substituting this ansatz into Eq.~\eqref{eq:dN/dt}, 
we find that $\kappa_N$ and $\kappa'_N$ are given by
 \begin{subequations}
 \label{eq:kappa_N}
   \begin{equation}
   \kappa_{N}^{2}+\frac{v}{D_{N}}\kappa_{N}-\frac{1}{l_{N}^{2}}=0
   \end{equation}
   \begin{equation}
   \kappa_{N}^{'2}-\frac{v}{D_{N}}\kappa'_{N}-\frac{1}{l_{N}^{2}}=0
   \end{equation}
  \end{subequations}
By matching the boundary conditions at $y-vt=0$, we find 
$g_{N}$ is given by
\begin{equation}
g_{N}=\frac{\kappa_{N}'}{\kappa_{N}+\kappa_{N}'}\label{eq:g_N}
\end{equation}

Essentially, this means that the shape of the concentration profile
depends on the relative magnitudes of $v$ and $v_{N}$. 
Physically, for $v>0$ and $v\gg v_{N}$,
the wavefront is traveling forward too fast for there to be much diffusion
from the producing region to the non-producing region, and so the
concentration at the front is low ($g_N\rightarrow1$); 
conversely, for $v<0$ and $|v|\gg v_{N}$,
the concentration at the front is high for the same reason.
(It is reminiscent of the Doppler effect, except that the Nodal signal
spreads by the diffusion equation rather  than the wave equation.)

There is in fact a symmetry relating solutions with $v>0$ and $v<0$, 
whereby the profiles of an advancing front of speed $v$ and a retreating front (reflected in $y$) of speed $-v$
add up to a uniform profile, because adding the production of an
advancing front and that of a retreating front simply obtains a uniform producing line. 
Furthermore, when $v=0$ the height of the front is simply $\frac{1}{2}N_{0}$,
and the stationary concentration profile simplifies to:
\begin{equation}
N(y)=\begin{cases}
N_{0}(1-\frac{1}{2}e^{y/l_{N}}) & ,\: y<0\\
\frac{1}{2}N_{0}e^{-y/l_{N}} & ,\: y>0
\end{cases}\label{eq:N_simple_step}
\end{equation}

(Notice that we set $y=0$ to be the position of the front.)
Since the coefficients in the threshold equations \eqref{eq:C_N}
and \eqref{eq:C_L} are in general different, the
Lefty wavefront is in general displaced by a distance $\Delta y$
from the Nodal one. We have

\begin{equation}
L(y-vt)=\begin{cases}
L_{0}[1-g_{L}e^{\kappa_{L}(y-vt+\Delta y)}] & ,\: y-vt<-\Delta y\\
L_{0}(1-g_{L})e^{-\kappa'_{L}(y-vt+\Delta y)} & ,\: y-vt>-\Delta y
\end{cases}\label{eq:L(y-vt)}
\end{equation}
where $\Delta y>0$ signifies that the Nodal front is ahead of the
Lefty front, and $\kappa_L$, $\kappa'_L$ and $g_L$ are defined analogously to 
Eqs.~\eqref{eq:kappa_N} and \eqref{eq:g_N}. 
Finally, to find $v$ and $\Delta y$, we
observe that $N=N_{0}(1-g_{N})$ at the Nodal front, and $L=L_{0}(1-g_{L})$
at the Lefty front, and substitute
 \eqref{eq:N-for-step-function} and \eqref{eq:L(y-vt)} into the threshold functions~\eqref{eq:Cij};
 assuming that $\Delta y>0$:
 \begin{subequations}
 \label{eq:v_Dy}
  \begin{equation}
  C_{NN}N_{0}(1-g_{N})-C_{NL}L_{0}(1-g_{L})e^{-\kappa'_{L}\Delta y}=1
  \end{equation}
  \begin{equation}
  C_{LN}N_{0}[1-g_{N}e^{-\kappa_{N}\Delta y}]-C_{LL}L_{0}(1-g_{L})=1
  \end{equation}
  \end{subequations}
from which we can solve (non-trivially) for $v$ and $\Delta y$.

\subsection{Nondimensionalization}
\label{sec:non-dim}

Certain combinations of parameter changes are trivial, in the
sense that the solutions look the same apart from rescalings
of the distance, time, or concentrations.
Since we are faced with the difficulty of exploring a large
parameter space, we wish to discover the minimum number of
nontrivial parameters.
To this end, we will scale all variables and parameters 
in the problem so as to make them dimensionless.
 
First, we introduce the scaled Nodal and
Lefty concentrations 
 \begin{subequations}
    \begin{equation}
     \tilde{N}(y)\equiv\frac{N}{N_{0}},
    \end{equation}
    \begin{equation}
    \tilde{L}(y)\equiv\frac{L}{L_{0}}, 
    \end{equation}
  \end{subequations}
so that $0\leq\tilde{N}(y),\:\tilde{L}(y)\leq 1$.

Correspondingly, 
$\tilde{C}_{NN}$, $\tilde{C}_{NL}$, $\tilde{C}_{LN}$ and $\tilde{C}_{LL}$
are simply the coefficients in the (scaled form of the) threshold functions 
\eqref{eq:C_N} and \eqref{eq:C_L}:
 \begin{subequations}
\label{eq:scaled-Cij}
\begin{eqnarray}
\tilde{C}_{NN} &\equiv& C_{NN}N_{0};\\
\tilde{C}_{NL} &\equiv& C_{NL}L_{0};\\
\tilde{C}_{LN} &\equiv& C_{LN}N_{0};\\
\tilde{C}_{LL} &\equiv& C_{LL}L_{0}.
\end{eqnarray}
\end{subequations}
For example, Nodal production turns on when $\tilde{C}_{NN}\tilde{N}-\tilde{C}_{NL}\tilde{L}>1$,
and similarly for $\tilde{L}$. 
The parameters \eqref{eq:scaled-Cij} are pertinent
even in a spatially uniform situation.

The remaining parameters relate to time or length scales.
We scale length and time such that the parameters for Nodal
become unity:
\begin{subequations}
\label{eq:scaled-l-v}
\begin{eqnarray}
\tilde{l}_L &\equiv& l_{L}/ l_{N}; \\
\tilde{v}_L &\equiv&    v_{L}/ v_{N}; \\
\tilde{v} &\equiv& v/v_{N}.
\end{eqnarray}
  \end{subequations}
The last two parameters are relevant (in a moving steady state)
if and only if $v\neq0$. 
Implicitly, Eqs.~\eqref{eq:scaled-l-v} also nondimensionalize
the time scale:
\begin{equation}
   \tilde{\tau}_L \equiv \tilde{l}_L/\tilde{v}_L
\end{equation}

In total, we have seven nontrivial parameters 
for the one-dimensional problem, defined in Eqs.~\eqref{eq:scaled-Cij}
and~\eqref{eq:scaled-l-v}. 
By rescaling space and time with $y\rightarrow\frac{y}{l_N}$ and 
$t\rightarrow\frac{t}{\tau_N}$ we find that Eqs.~\eqref{eq:d/dt} become
\begin{subequations}
\label{eq:tilded/dt}
\begin{equation}
\frac{\partial \tilde{N}(y,t)}{\partial t}=\frac{\partial^{2}\tilde{N}(y,t)}{\partial y^{2}}-\tilde{N}(y,t)+\tilde{s}_{N}(y,t)
\label{eq:dtildeN/dt}
\end{equation}
\begin{equation}
\frac{\partial \tilde{L}(y,t)}{\partial t}=\tilde{l}_L\tilde{v}_L\frac{\partial^{2}\tilde{L}(y,t)}{\partial y^{2}}-
\frac{1}{\tilde{\tau}_{L}}\tilde{L}(y,t)+\tilde{s}_{L}(y,t)
\label{eq:dtildeL/dt}
\end{equation}
\end{subequations}
where $0\leq\tilde{s}_N(y,t)\leq1$ and $0\leq\tilde{s}_L(y,t)\leq\frac{1}{\tau_L}$.

\subsection{The $(N,L)$ plane: zero dimensional case}
\label{sec:N-L-plane}

\begin{figure}[t!]\vspace*{3pt}
\centering{\includegraphics[width=3.2in]{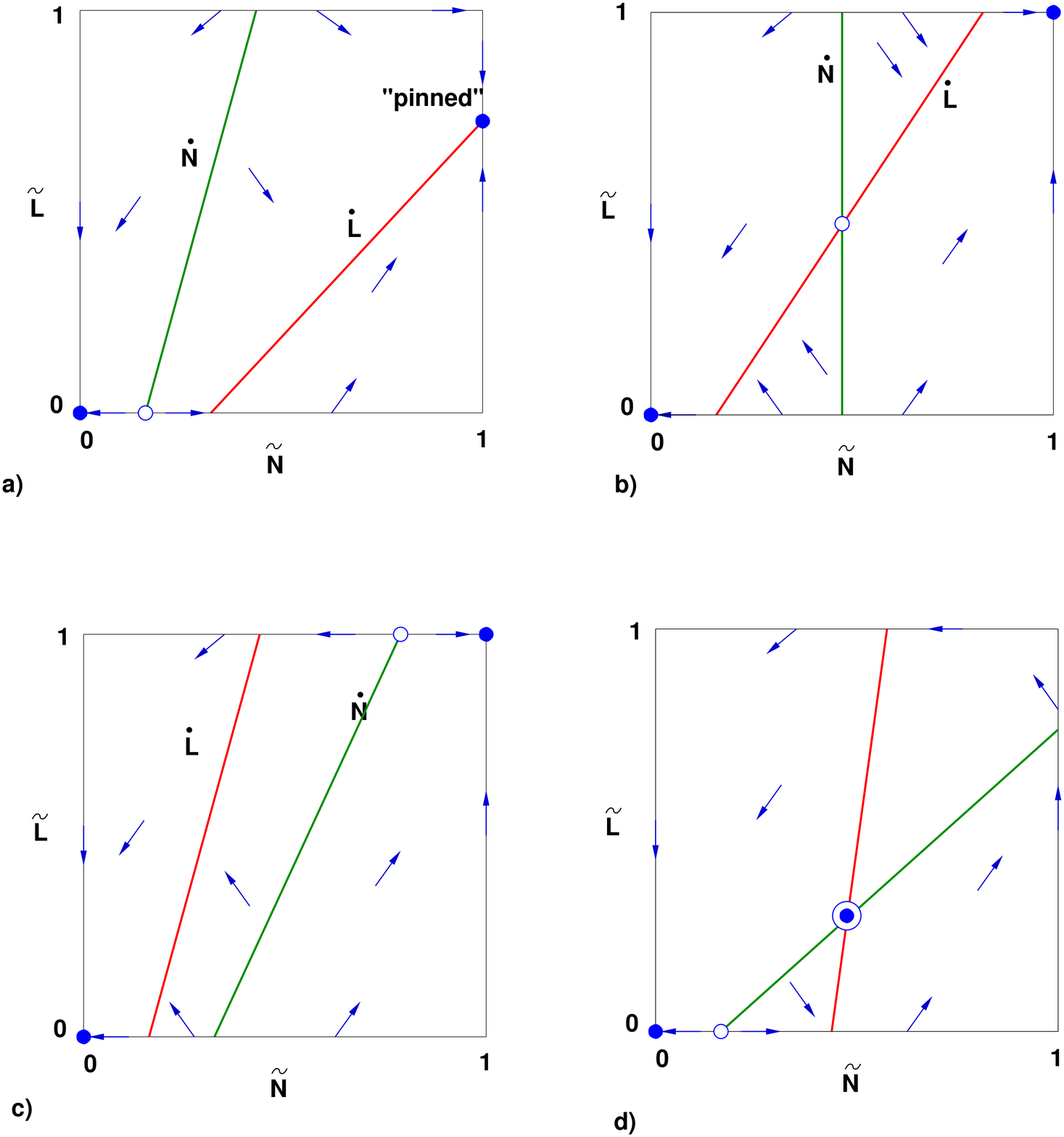}}\vspace*{-5.5pt}
\caption{
\label{fig:N-L-cases}
Possible behaviors on the $(\tilde{N},\tilde{L})$ phase plane. 
Green and red lines represent
$\dot{\tilde{N}}(\tilde{N},\tilde{L})=0$ 
and $\dot{\tilde{L}}(\tilde{N},\tilde{L})=0$,
respectively.
Stable (unstable) fixed points are marked by
closed (open) circles; the circle in (d) with a dot 
is a fixed point that may either be stable or
may become unstable to a cycle, as elaborated
in Sec.~\ref{sec:oscillating}.
The topology is classified into cases (a),(b),(c),
and (d), following Sec.~\ref{sec:N-L-plane}, 
according to whether the $\dot{\tilde{N}}$ isocline
is to the right or the left of the $\dot{\tilde{L}}$
isocline, or how they cross (if they do).
Each case has several subcases according to where the
respective isoclines interect the bounding square.
Cases (a) and (c) are compatible with
Eqs.~\eqref{eq:phi-to-linear}, which requires
that both isoclines have the same vertical intercept;
cases (a),(b), or (c) might be compatible with
the condition $C_{NL}=0$ adopted in Sec.~\ref{sec:results},
which required that the $\dot{\tilde{N}}$ isocline is
vertical.
Either of cases (a) or (b) is consistent with 
a fixed point at $0<\tilde{L}_*<1$, associated with
``pinned'' intervals throughout which $(N,L)$  is
exactly on the $\dot\tilde{L}=0$ isocline (Sec.~\ref{sec:pinned}).
}
\vspace*{-8pt}
\end{figure}

In Sec.~\ref{sec:traveling-wavefront}, asymptotically
in either direction both $N$ and $L$ approach constant,
uniform values. 
Therefore, in order to classify possible
wavefront solutions, it is helpful first to classify
all possible uniform solutions.
Fig.~\ref{fig:N-L-cases} shows the $(\tilde{N},\tilde{L})$ phase plane with lines representing 
the $\dot{\tilde{N}}=0$ and $\dot{\tilde{L}}=0$ isoclines. The most important feature of these diagrams
is the relation of the two isoclines to each other and to the bounding lines: this determines the 
possible steady states. 

There are 4 qualitative cases for the relation of $\dot{\tilde{N}}$ to $\dot{\tilde{L}}$:
\begin{enumerate}
\item $N$-isocline is to left of $L$-isocline
\item $L$-isocline crosses $N$-isocline left to right
\item $L$-isocline is to left of $N$-isocline
\item $N$-isocline crosses $L$-isocline left to right.
\end{enumerate}
(There are other mathematical cases
in which one of the isoclines does not pass
through the square at all, but that is 
obviously to be discarded, since there would be
no way ever to produce the corresponding signal.)

These phase planes are generalizations of the wild-type scenario:
as noted after Eq.~\eqref{eq:phi-to-linear}, if we justify
the linear threshold equations \eqref{eq:Cij} from the
single-binding site receptor behavior \eqref{eq:phi},
we necessarily find $\tilde{C}_{NL}=\tilde{C}_{LL}$;
the geometrical expresion of this on the $(\tilde{N},\tilde{L})$ plane
is that the two isoclines cross on the negative $L$ axis,
hence only cases (a) and (c) could be realized.
However, our purpose here is to study the general
behavior of these equations 
that could arise in any embryonic system with propagating Nodal and Lefty fronts
(since this is conserved in all vertebrates),
and perhaps even mathematically equivalent equations having
a quite different biological interpretation.
In general, binding of Nodal and Lefty might be cooperative
at the receptor; furthermore, gene regulation
further downstream might be cooperative. Thus, the general form of the isoclines
is nonlinear and we expect that all these topologies
are imaginable.

\LATER{\CLH{ (AND WHAT DID REF.~\cite{nakamura-hamada} SAY
ABOUT THE ISOCLINES?  I BELIEVE THEY WORRIED ABOUT THESE CASES.)}}

There is always a stable fixed point at $\tilde{N}=\tilde{L}=0$;
this is the only fixed point in case (iii), which is thus trivial.  
If the $\tilde{N}$-isocline and the $\tilde{L}$-isocline both intersect the
upper border, that is if $\tilde{C}_{NN}-\tilde{C}_{NL}>1$
and $\tilde{C}_{LN}-\tilde{C}_{LL}>1$, we have a stable
fixed point with both $N$ and $L$ saturated, which underlies
the basic wavefront behavior of Sec.~\ref{sec:traveling-wavefront}
above.

However, if the $\tilde{L}$-isocline intersects the right edge,
to the right of the $\tilde{N}$-isocline, i.e. when $\tilde{C}_{LN}-\tilde{C}_{LL}<1$ as {\it may} 
happen in either case (i) or case (ii),
then the stable fixed point at $\tilde{N}=1$ exists at a less than saturated value, 
which we will define $\tilde{L}_{*}$:
\begin{equation}
\tilde{L}_{*}\equiv\frac{\tilde{C}_{LN}-1}{\tilde{C}_{LL}}\label{eq:L*}
\end{equation}
It is not obvious that $\tilde{L}$ being {\it pinned} at $\tilde{L}_*<1$ exists in the case with spatial
variation, but we will in fact show in Sec.~\ref{sec:pinned} that analogous ``pinned'' intervals
arise in 1D traveling wave solutions.

Finally, in case (iv), there is a fixed point at which
{\it both} $\tilde{N}$ and $\tilde{L}$ are less than saturated; this may
be stable, but also may be unstable to periodic oscillations
around the fixed point, as elaborated in Sec.~\ref{sec:oscillating}.

\section{Results: steady-state fronts}
\label{sec:results}

To explore the large parameter space of the 1D model, we will consider 
special cases; the aim is a taxonomy of the possible
behaviors of the model, with the hope that any general case
will be qualitatively similar to one of these behaviors
seen in the special cases. In particular, we concern ourselves
with the steady state limiting behaviors, after all transients have died out.
Any nonzero Nodal and Lefty wavefronts will then be traveling at identical
speed $v$.

\subsection{$\tilde{C}_{NL}=0$ and $v=0$}

For the largest part of our story, we consider the case
that $\tilde{C}_{NL}=0$, that is, Nodal is unaffected by Lefty
(i.e. Nodal is an autonomous variable).
We first solve for 
the behaviors of the one-component system in which Nodal 
activates Nodal, which already has moving front solutions;
then, we treat the Nodal concentration $N(y,t)$ as if it
were externally imposed and solve for another one-component
problem representing the Lefty concentration field.
The reasons for choosing what seems to be a major simplification are
(i) the reverse approximation, in which Lefty is autonomous, 
has only the trivial solution with Lefty off (since Lefty only inhibits);
(ii) experimentally, the Nodal front leads and the Lefty front
follows; thus, it is plausible that the inhibition from Lefty has
no important effect on the Nodal front (serving only to prevent
the initiation of a front on the other side);
(iii) in Section~\ref{sec:N-L-plane} we see that only the 
topology of the $(N,L)$ plot
really matters for the dependnce on coefficients $C_{ij}$.

The second simplification we make is to set $v=0$. As mentioned, the Nodal and
Lefty fronts are traveling at equal speed in the steady state, and are thus stationary in
the comoving frame. In fact, the only difference between the $v=0$ and $v\neq0$ cases
is the steepness of their exponential profiles at the front. This does {\it not} qualitatively affect
the results we present below.

With $\tilde{C}_{NL}=0$ and $v=0$, the Nodal concentration takes the form given by
\eqref{eq:N_simple_step}, or in dimensionless form,
\begin{equation}
\tilde{N}(y)=\begin{cases}
1-\frac{1}{2}e^{y} & ,\: y<0\\
\frac{1}{2}e^{-y} & ,\: y>0
\end{cases}\label{eq:N_simple_step_dimless}
\end{equation}
and our 7-dimensional parameter space reduces to a 
3-dimensional one, with parameters $\tilde{l}_L$, $\tilde{C}_{LL}$, and $\tilde{L}_{*}$ 
(which is more convenient for our purposes than $\tilde{C}_{LN}$). Our goal is to classify
all the possible types of behavior of the Lefty profile as these parameters are varied.
Note that $\tilde{C}_{NN}$ and $\tilde{C}_{LN}$ are always greater than 1, or
else N and L will never have their production turned on. Also, let the position where Lefty 
production turns on be $y=y_{L}$; it may be on either side of the independent Nodal front $y=0$.

\subsection{``Pinned'' intervals}
\label{sec:pinned}

It soon became apparent that the behavior of $L$ did not only consist of intervals of full production
$s_{L}=s_{0L}$, where $C_{L}(y)>1$, and zero production, where $C_{L}(y)<1$, 
but also intervals where the threshold
function is \emph{pinned} at $C_{L}(y)=1$. The clearest example of this is to consider what happens
when $\tilde{L}_{*}<1$ as $y\rightarrow-\infty$, where our background Nodal profile goes to a limit of 
$\tilde{N}\rightarrow1$.
Since we want $L$ to be produced somewhere, $\tilde{C}_{LN}\equiv C_{LN}N_{0}>1$ necessarily
(this is \eqref{eq:C_L} with $L=0$). As $y\rightarrow-\infty$ and $\tilde{N}\rightarrow1$, 
$L(y)=0$ is mathematically inconsistent, and so Lefty production must turn on. Now, by definition $\tilde{L}_{*}$ is the amount of $L$ that turns off its own production when $N=N_{0}$,
so $\tilde{L}_{*}<1$ means that before $\tilde{L}\rightarrow1$, Lefty will turn off its own production again. This is a ``paradox'' in which Lefty production cannot be fully on or fully off, and it is only resolved if we let 
the threshold function be \emph{pinned} at 1, i.e. $C_L(y\rightarrow-\infty)\equiv1$, such that $0<\tilde{s}_{L}<1$. Constraining $C_L(y)$ means that $\tilde{L}(y)$ is completely determined by 
$\tilde{N}(y)$ in this \emph{pinned} interval:

\begin{equation}
\tilde{L}(y) \equiv \frac{1}
{\tilde{C}_{LL}}\label{eq:L_pinned}
\bigg(\tilde{C}_{LN}\tilde{N}(y)-1\bigg)
\end{equation}

Now, we explicitly work out the possible behaviors of $L$ when $\tilde{L}_{*}$,
$\tilde{l}_L$ and $\tilde{C}_{LL}$ are varied. $L$ will be described in terms of the types of production
occurring along the line, from $-\infty$ to $\infty$. Full production is denoted ``1'', zero production
denoted ``0'', and the pinned interval denoted ``p''. For example, the background Nodal profile is
characterized as \{1,0\}, which is shorthand for an interval of full production adjoining one of zero
production. In the following, we split up the cases into $\tilde{L}_{*}<1$ and $\tilde{L}_{*}>1$.
\CLHdone{I would make a point somewhere that, for these behaviors, the
most important parameter becomes the ratio of length scales,
$\tilde{l}_L$. The value of $\tilde{v}_L$ does not enter at all.}
We find that the most important parameter is the ratio of length scales, $\tilde{l}_L$, which determines
how various pinned intervals arise in the behavior of Lefty. 
Note that the value of $\tilde{v}_L$ does not enter at all.

\subsection{Classification of 1D model}

\subsubsection{Case $\tilde{L}_{*}<1$: \{p,0\} and \{p,1,0\}}
\label{sec:pinned-lt1}

\begin{figure}[t!]\vspace*{3pt}
\begin{center}$
\begin{array}{cc}
 \includegraphics[width=3.3in]{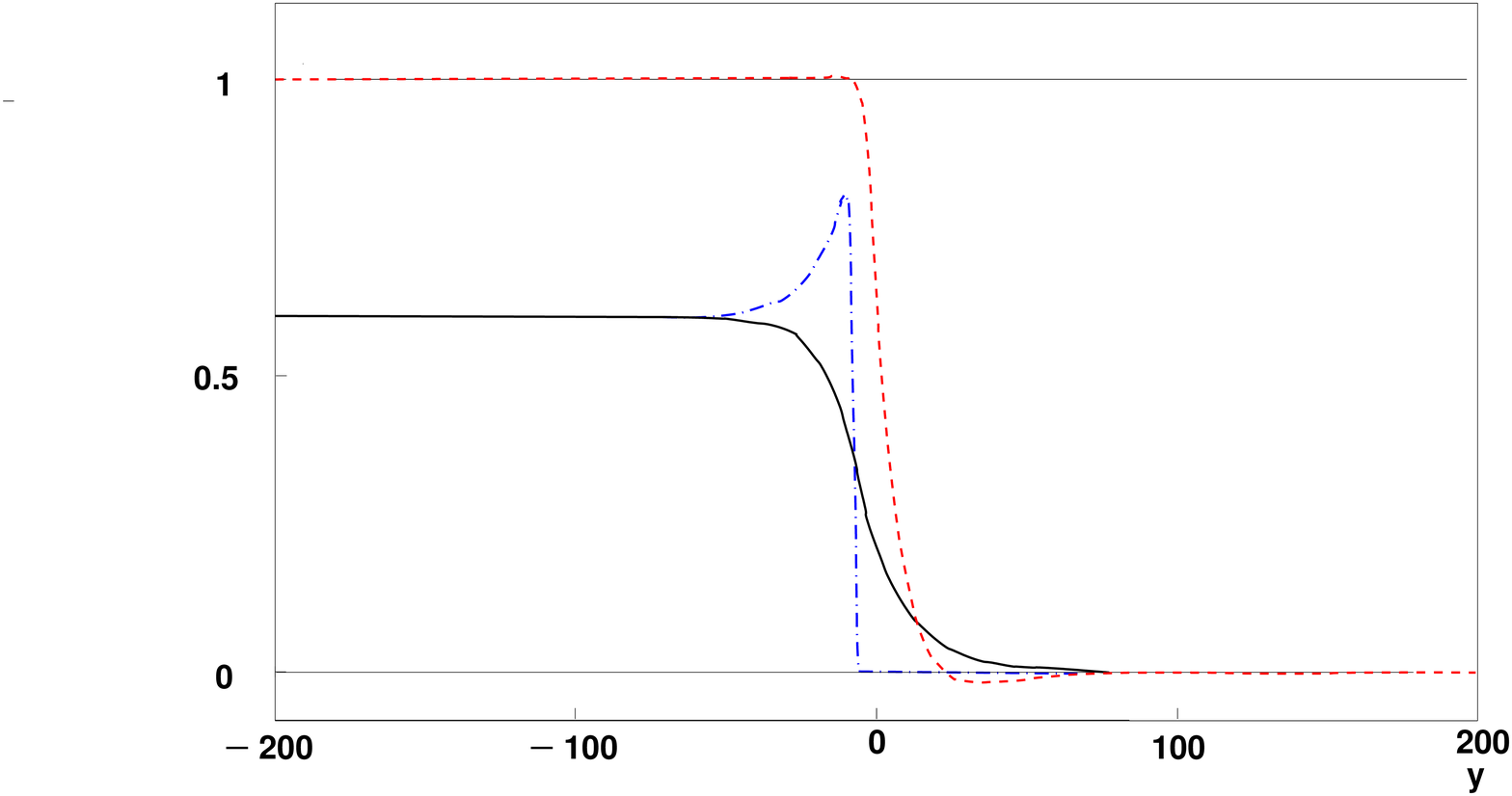} \\
 \includegraphics[width=3.3in]{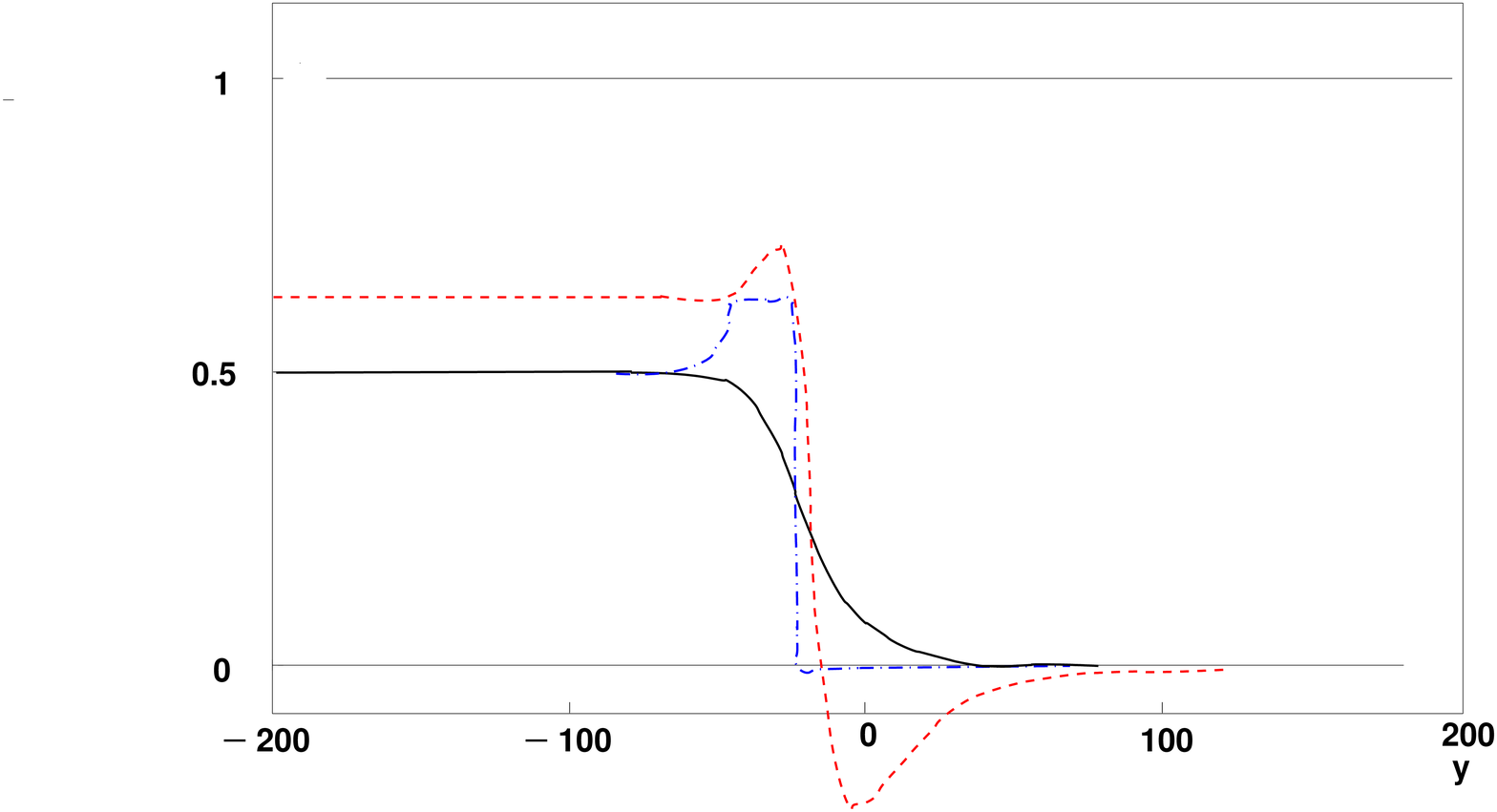}
\end{array}$
\end{center}\vspace*{-5.5pt}
\caption{
\label{fig:pin1}
Form of wavefront when $\tilde{L}*<1$
(section ~\ref{sec:pinned-lt1}).
Top: \{p,0\}. Bottom: \{p,1,0\}.
Graphs are $C_L(y)$ (red dashed line), $\tilde{s}_L(y)$ (blue dot-dash line) 
and $\tilde{L}(y)$ (black solid line); the threshold is unity.
Parameter values:  $\tilde{L}*=0.8$, $\tilde{l}_L=1$ (top), $\tilde{l}_L=1.5$ (bottom)
}\vspace*{-8pt}
\end{figure}

As already shown, there is a pinned interval extending to $y\rightarrow-\infty$, where $L$ is completely
determined. The production $s_L(y)$ is also completely determined, and is derived as follows. 
Substituting \eqref{eq:L_pinned} back into (\ref{eq:dL/dt}):
\[
\frac{D_{L}}{\tilde{C}_{LL}}\frac{d^2\tilde{N}(y)}{dy^2}-\frac{1}{\tau_{L}}\tilde{L}+\tilde{s}_{L}(y)=0
\]
Now using (\ref{eq:dN/dt}) to eliminate $d^2\tilde{N}(y)/dy^2$, Lefty production
in the pinned interval is given by:

\begin{equation}
\tilde{s}_{L}(y)=\frac{1}{\tilde{C}_{LL}}(\tilde{C}_{LN}[(1-\tilde{l}_L^{2})\tilde{N}(y)+\tilde{l}_L^{2}\tilde{s}_{N}(y)]-1)
\label{eq:production_pinned}
\end{equation}
where $\tilde{s}_{N}(y)=\begin{cases}
0 & , y>0\\
1 & , y<0
\end{cases}$.

The point of finding this expression is that as $y\rightarrow-\infty$, 
we see that $\tilde{s}_{L}\rightarrow\tilde{L}_{*}<1$. This simply means that as $\tilde{L}_{*}$
goes to zero, so does the asymptotic production of Lefty.

The values of $\tilde{l}_{L}$, $\tilde{C}_{LN}$ or $\tilde{C}_{LL}$
in (\ref{eq:production_pinned}) are free to vary so long as $\tilde{L}$ in \eqref{eq:L_pinned} is between 0 and 1.
On the other hand, the location of the front separating the pinned interval and the region
of zero production must be determined from matching both $\tilde{L}$ and $d\tilde{L}/dy$
on either side of the boundary.
It is possible to end up with a value of $y_{L}$ where $\tilde{s}_{L}$
lies outside {[}0,1{]}. Indeed, checking when when this occurs using
(\ref{eq:production_pinned}) demonstrates that \{p,0\} is inconsistent
when 
\begin{equation}
\tilde{L}_{*}\tilde{l}_L>1
\end{equation}
\LATER{I have some inequalities for whether $y_L<0$ or $>0$, and I can show why $s_L$
cannot lie outside $[0,1]$ for the other cases.}
Here, there are three regions instead: pinned, full production,
and no production (in short, \{p,1,0\}). As $\tilde{L}_{*}\tilde{l}_L$
increases beyond 1, the length of the fully producing interval increases.
Simulations and solving for the boundary conditions validate this (see figure~\ref{fig:pin1}).

\subsubsection{Case $\tilde{L}_{*}>1$: \{1,0\} and \{1,p,0\}}
\label{sec:pinned-gt1}

\begin{figure}[t!]\vspace*{3pt}
\begin{center}$
\begin{array}{cc}
 \includegraphics[width=3.3in]{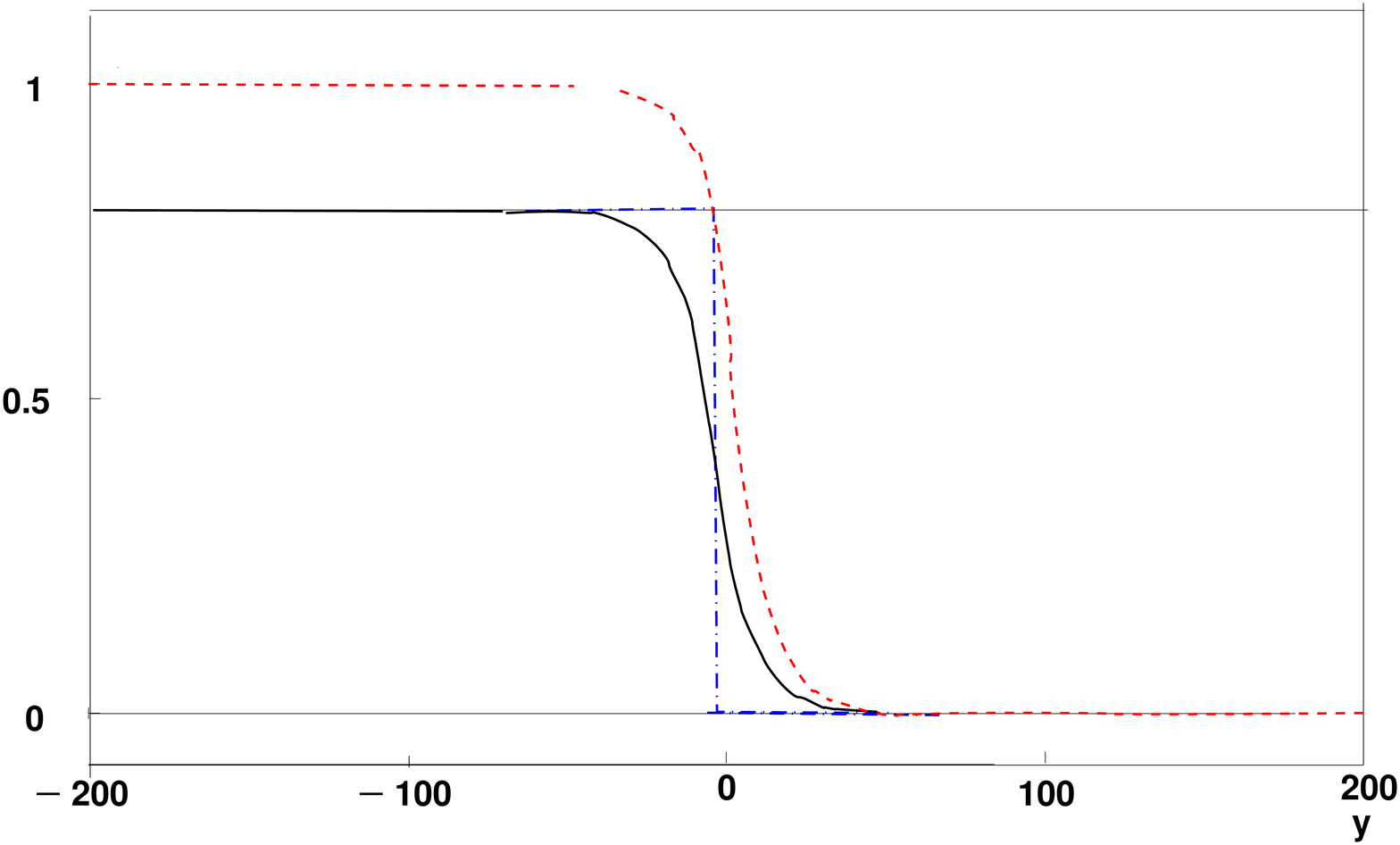} \\
 \includegraphics[width=3.3in]{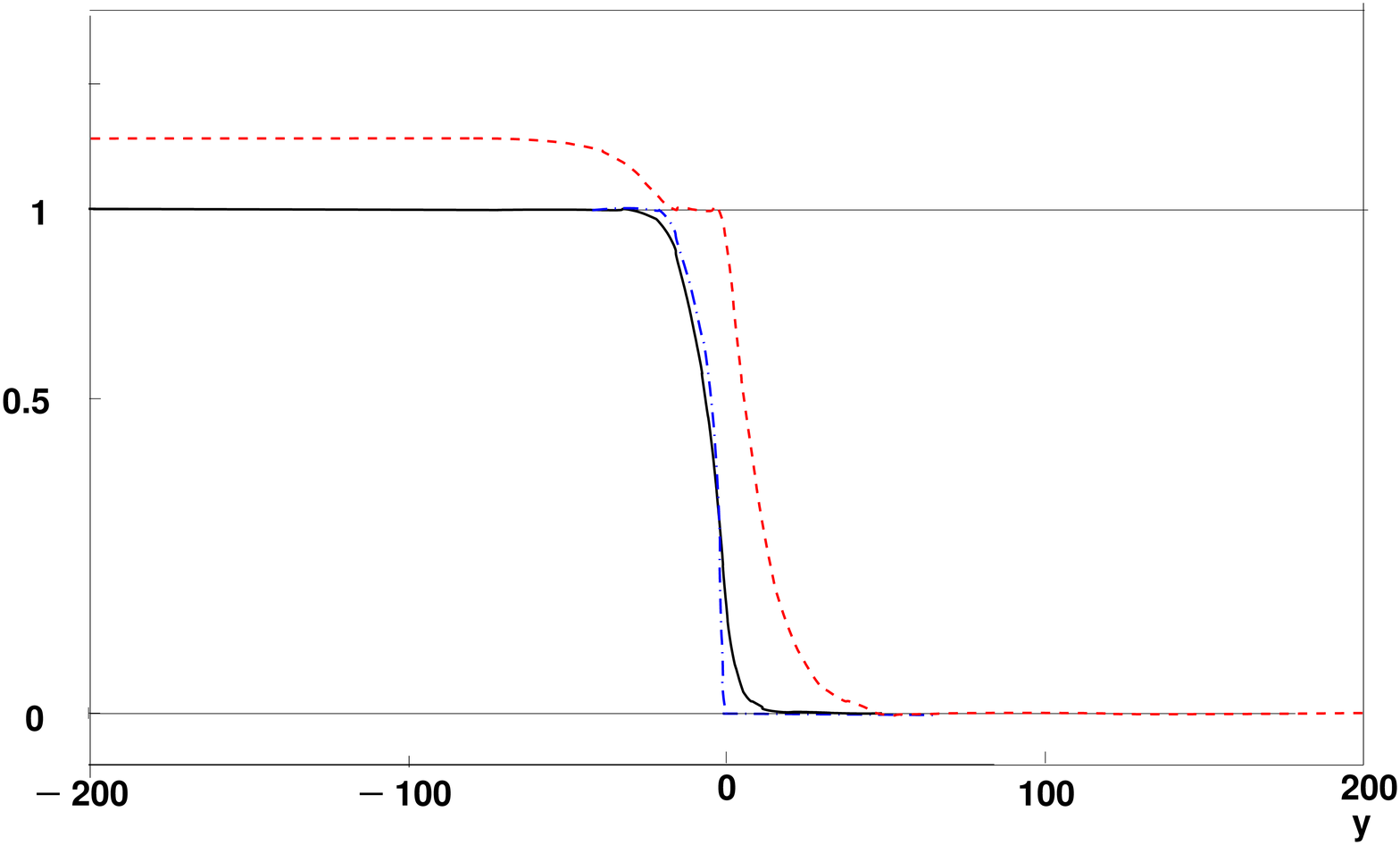}
\end{array}$
\end{center}\vspace*{-5.5pt}
\caption{
\label{fig:pin2}
Form of wavefront when $\tilde{L}*>1$
(section ~\ref{sec:pinned-gt1}).
Top: \{1,0\}. Bottom: \{1,p,0\}.
The curves $C_L(y)$ (red dashed), $\tilde{s}_L(y)$ (blue dot-dash) and 
$\tilde{L}(y)$ (black solid) as in Figure \ref{fig:pin1}.
\LATER{The threshold is the red dotted line.}
Parameter values:  $\tilde{L}_*=1.5$, $\tilde{l}_L=1$ (top), $\tilde{l}_L=0.4$ (bottom)
}\vspace*{-8pt}
\end{figure}

The traveling wavefront given by (\ref{eq:L(y-vt)}) is still not guaranteed when $\tilde{L}_{*}>1$.
We can see this by considering the limit at which $\tilde{l}_L\ll1$: at the front of Lefty production,
 $\tilde{L}$ goes from 1 to 0 with a background of nearly constant $N$. Since
$L$ inhibits its own production, the apparent paradox is that $dC_{L}(y)/dy>0$,
i.e. the regions of zero and full production are reversed! Simulations
show that as $\tilde{l}_L$ is decreased and the slope of $L$ made steeper,
once the slope exceeds that of the ``pinned'' Lefty (\ref{eq:L_pinned}),
\{1,0\} will transition to \{1,p,0\}. Basically, requiring that $dC_{L}(y)/dy<0$
at the front is equivalent to this condition.

To write out explicit conditions for entering the \{1,p,0\} state, we should consider both cases 
$y_{L}>0$ and $y_{L}<0$, for which $\tilde{N}$ takes
different forms given in \eqref{eq:N_simple_step_dimless}. Thus, the condition for \{1,p,0\} is that $C_L(y)$ crosses the threshold
in the other direction: $dC_{L}(y_{L})/dy>0$, giving:
\begin{subequations}
\begin{eqnarray}
y_{L}&>0:\;\tilde{l}_L^{-1}>&1+\frac{2}{\tilde{C}_{LL}} \label{eq:{1,pin,0}-1} ;\\
y_{L}&<0:\;\tilde{l}_L^{-1}>& 2\tilde{L}_* - 1. \label{eq:{1,pin,0}-1-1}
\end{eqnarray}
\end{subequations}
(Note that both conditions require $\tilde{l}_L<1$, 
i.e. the Nodal length scale is greater than Lefty, as argued above
in the limiting case.)
If the inequalities are in the opposite direction, then we are back to the traveling
wavefront given by (\ref{eq:L(y-vt)}), or in our notation {1,0}. 
Here we can solve for the Lefty concentration profile fully: using 
\eqref{eq:v_Dy} we find that $\Delta y>0$ when
\begin{equation}
\frac{1}{2}\tilde{C}_{LN} - \frac{1}{2}\tilde{C}_{LL} < 1.
\end{equation}

\begin{table}
\begin{tabular}{|c|c|}
\hline 
Interval patterns & Parameter conditions\tabularnewline
\hline
\hline
$\{1,0\}$ & $\tilde{L}_{*}>1,\:{dC_{L}(y_{L})}/{dy}<0$\tabularnewline
\hline 
$\{1,p,0\}$ & $\tilde{L}_{*}>1,\:{dC_{L}(y_{L})}/{dy}>0$\tabularnewline
\hline 
$\{p,0\}$ & $\tilde{L}_{*}<1,\:\tilde{l}_{L}^{-1}>\tilde{L}_{*}$\tabularnewline
\hline 
$\{p,1,0\}$ & $\tilde{L}_{*}<1,\:\tilde{l}_{L}^{-1}<\tilde{L}_{*}$\tabularnewline
\hline 
\end{tabular}

\caption{Conditions for possible Lefty interval patterns, given
a Nodal wavefront for $\tilde{C}_{NL}=0$ and $v=0$. Note that there are
3 independent parameters: $\tilde{l}_{L}$, $\tilde{C}_{LL}$, and
$\tilde{L}_*$ (or $\tilde{C}_{LN}$)}

\end{table}

\section{Solutions oscillating in time}
\label{sec:oscillating}

In the previous section(s), we explored the limiting case of an autonomous, 
Lefty-independent Nodal production. 
We now address new phenomena which are possible when Nodal is
significantly affected by Lefty.
The most striking of these is the possibility of a time-oscillating
solution, which we will study in the special case that all 
concentrations are uniform in space.  
The experimentally pertinent motivation is whether it is possible to
generate traveling solutions,  consisting of pulses of Nodal and Lefty,
periodic in space and in time; one expects such solutions to
be possible only if uniform oscillations are possible in the model.
\LATER{Why?}

By examining the geometry of the $(N,L)$ plane [Figure~\ref{fig:N-L}],
it can be seen that only case (d) has the possibility of oscillations;
this is shown with more detail in Fig.~\ref{fig:N-L-osc}(b).
Note that the fixed point in the center, since it has $N_F<N_0$ and $L_F<L_0$,
corresponds to production rates less than saturation ($s_N<s_{N0}$ and $s_L<s_{L0}$),
so we are in a ``pinned'' regime for {\it both} $N$ and $L$, which is self-consistent
with being on the isoclines for both. 

\subsection{Parts of each cycle}

\begin{figure}[t!]\vspace*{3pt}
 \centering{\includegraphics[width=2.6in]{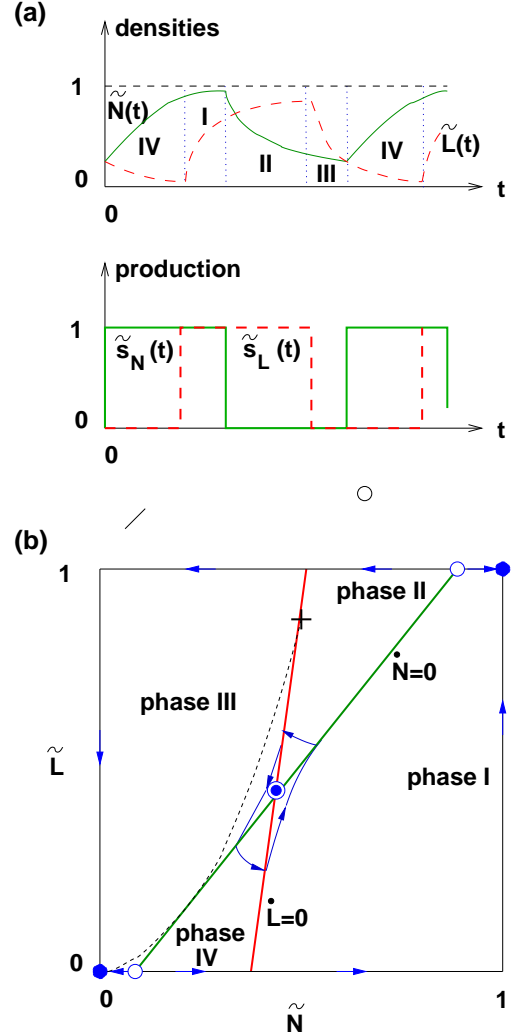}}\vspace*{-5.5pt}
\caption{
\label{fig:N-L-osc}
(a). Dynamics of the concentrations $\tilde{N}(t)$ and $\tilde{L}(t)$ of Nodal and Lefty, and
their production rates $s_N(t)$ and $s_L(t)$, in the case of periodic oscillation
(schematic).
Phases of the cycle are labeled I,II, III, and IV.
(b). Phase plane of the concentrations. 
Heavy lines indicate the sign change of $\dot{\tilde{N}}(\tilde{N},\tilde{L})$ (green) and of 
$\dot{\tilde{L}}(\tilde{N},\tilde{L})$ (red). 
A trajectory (blue) is shown, which (initially) is spiraling out from an
unstable fixed point $(\tilde{N}_*,\tilde{L}_*)$.
Filled and open circles are stable and unstable fixed points.
The trajectory from the heavy cross is tangent to the
next isocline; any trajectory starting higher goes to $(0,0)$.
}\vspace*{-8pt}
\end{figure}

The period of an oscillation must consist of four phases
[as shown in Fig.~\ref{fig:N-L-osc}(a)]:
\begin{itemize}
\item[(I)]
Nodal and Lefty both on; Lefty grows faster than 
Nodal, until Nodal turns off;
\item[(II)]
Nodal off and Lefty on; Nodal decays, until Lefty turns off;
\item[(III)]
Nodal and Lefty both off;
Lefty decays faster than Nodal, and they remaining Nodal 
that has not yet decayed is sufficient to turn on Nodal again;
\item[(IV)]
Nodal on and Lefty off;
Nodal grows, until it is sufficient to turn Lefty on too
(and back to phase 1).
\end{itemize}
Evidently, a key condition is on the degradation times,
which govern the growth rates of Nodal or Lefty
: we need
   \begin{equation}
          \ttauL \equiv \tau_L/\tau_N < 1.
   \label{eq:nu-tauN-tauL}
   \end{equation}
(Of course, the actual magnitude of the degradation times
simply sets an overall time scale.)

We can describe the history with a series of times $t_i$, i=1,2,..., 
at each of which one of the signals turns on or off.
Between these times, the production rates are constant (either zero
or saturated), and the differential equation is solved by a simple
exponential:
   \begin{equation}
      \tilde{N}(t) = \begin{cases}
                \tilde{N}(t_i) e^{-(t-t_i)/\tau_N} &, \: s_N=0\\
                1 -[1-\tilde{N}(t_i)] e^{-(t-t_i)/\tau_N} &, \: s_N=s_{N0},
             \end{cases}
   \label{eq:N-t-constant-s}
   \end{equation}
and similarly for $\tilde{L}(t)$.

To visualize the dynamics, it is convenient to draw the trajectory
of $(\tilde{N}(t),\tilde{L}(t))$ in the $(\tilde{N},\tilde{L})$ plane; in our equations,
the time derivatives $\dot{\tilde{N}}\equiv d\tilde{N}/dt$ and $\dot{\tilde{L}}\equiv d\tilde{L}/dt$
are fuctions only of $(\tilde{N},\tilde{L})$.
The lines $\tilde{C}_N(\tilde{N},\tilde{L})=0$ and $\tilde{C}_L(\tilde{N},\tilde{L})=0$ on this plot are the 
{\it isoclines}, meaning the places where (respectively) 
$\dot{\tilde{N}}$ and $\dot{\tilde{L}}$ change sign.
(Isoclines were also used in the analysis of Ref.~\cite{nakamura-hamada}.)
Each time the trajectory intersects an isocline is one of the times 
$t_i$; let the concentrations at these times be
$\tilde{N}_i\equiv\tilde{N}(t_i)$,
$\tilde{L}_i\equiv\tilde{L}(t_i)$.
We can find the trajectory curve, and thus find solutions
of the dynamical equations, by eliminating time and considering
only the discrete map $(\tilde{N}_i,\tilde{L}_i)\to (\tilde{N}_{i+1},\tilde{L}_{i+1})$.
We next find the functional form of this map.

Consider phase III of the cycle, during which 
[according to \eqref{eq:N-t-constant-s}]
$\tilde{N}(t) = \tilde{N}_i e^{-(t-t_i)/\tau_N}$ and
$\tilde{L}(t) = \tilde{L}_i e^{-(t-t_i)/\tau_L}$.
Eliminating time, we get 
\begin{subequations}
\label{eq:N-L-phases}
\begin{equation}
\frac{\tilde{L}}{\tilde{L}_i} = \Biggl[\frac{\tilde{N}}{\tilde{N}_i}\Biggr]^{1/\ttauL}.
\label{eq:N-L-phaseIII}
\end{equation}
The point $(\tilde{N}_{i+1},\tilde{L}_{i+1})$ is the intersection
of the curve \eqref{eq:N-L-phaseIII} with the
isocline $\tilde{C}_N(\tilde{N},\tilde{L})=0$. In phase IV, the trajectory would be
\begin{equation}
\frac{\tilde{L}}{\tilde{L}_i} = \Biggl[\frac{1-\tilde{N}}{1-\tilde{N}_i}\Biggr]^{1/\ttauL},
\label{eq:N-L-phaseIV}
\end{equation}
\end{subequations}
and similarly in the other four phases of the cycle.
Notice that the curve can possibly intersect the isocline
in phases I or III only if $\ttauL < 1$. (For example,
whehn $\ttauL=1$ the trajectory in each phase is a straight 
line to one corner of the square.)

To give the cycling behavior, it is necessary 
(but not sufficient) that the
isoclines intersect as shown in Figure~\ref{fig:N-L-cases}(b),
which depends on two inequalities.  
The $\dot{\tilde{N}}$ and $\dot{\tilde{L}}$ 
isoclines' respective intercepts at $\tilde{L}=0$ satisfy
$1/\tilde{C}_{NN} < 1/\tilde{C}_{LN}$; their intercepts  at
$\tilde{L}=1$ must be in the reverse order,
$(1+\tilde{C}_{NL})/\tilde{C}_{NN} > (1+C_{LL})/\tilde{C}_{LN}$. 
The two inequalities can be written together as
\begin{equation}
   \frac{\tilde{C}_{LL}+1}{\tilde{C}_{NL}+1}  <
    \frac{\tilde{C}_{LN}} {\tilde{C}_{NN}} < 1
   \label{eq:osc-fp-cond}
\end{equation}
This tends to be satisfied 
When conditions \eqref{eq:osc-fp-cond} are satisfied, 
the isoclines cross  at a fixed point $(\tilde{N}_*,\tilde{L}_*)$ 
where
\begin{subequations}
\begin{eqnarray}
\tilde{N}_* &\equiv& \frac {\tilde{C}_{NL}-\tilde{C}_{LL}}
{\tilde{C}_{LN} \tilde{C}_{NL}- \tilde{C}_{NN}\tilde{C}_{LL}} \\
\tilde{L}_* &\equiv& \frac {\tilde{C}_{NN}-\tilde{C}_{LN}}
{\tilde{C}_{LN} \tilde{C}_{NL}- \tilde{C}_{NN}\tilde{C}_{LL}}.
\end{eqnarray}
\end{subequations}

\subsection{Stability conditions}

The trajectory always tends to spiral around 
$(\tilde{N}_*,\tilde{L}_*)$.
as illustrated in Figure~\ref{fig:N-L-osc}(b).
However, there are different possibile limiting behaviors
depending on whether the trajectory spirals inwards (stable fixed point)
or outwards, and on where it ends up.
There is always a stable fixed point at 
$(\tilde{N},\tilde{L})=(0,0)$ [the empty system]
and provided the isoclines intercept the upper edge of 
$(N,L)$ space [as in Figure ~\ref{fig:N-L-osc}(b)],
there is also a stable fixed point at 
$(\tilde{N},\tilde{L})=(1,1)$ [both Nodal and Lefty turned on and saturated].
Some of the possibilities:
\begin{itemize}
\item[(1)] The fixed point is stable; there is
one unstable cycle, such that a trajectory starting 
inside it tends to $(\tilde{N}_*,\tilde{L}_*)$ and a trajectory
starting outside it tends to $(0,0)$ or $(1,1)$.
\item[(2)] There are no cycles (stable or unstable);
the fixed point is unstable, and spirals out to 
$(0,0)$ or $(1,1)$.
\item[(3)] The fixed point is unstable, and
spirals out to a stable cycle (beyond which is an 
untable cycle, as in case (1)); this is the case
of interest to us.
\end{itemize}

To evaluate the stability, we must linearize 
near the fixed point in terms of 
$\delta \tilde{N}\equiv \tilde{N}-\tilde{N}_*$,
$\delta \tilde{L}\equiv \tilde{L}-\tilde{L}_*$.
The trajectory (e.g. Eqs.~ \eqref{eq:N-L-phaseIII}
or \eqref{eq:N-L-phaseIV})
becomes $\delta \tilde{L}=m_{\rm ph}\delta \tilde{N}$,
 also $m_{\rm ph}$
is the slope in each respective phase of the cycle,
${\rm ph}\to$ I,II,III, or IV.
Thus,
\begin{subequations}
\begin{equation}
m_\phaseIII= \frac{1}{\ttauL} \frac{\tilde{L}_*}{\tilde{N}_*};
\end{equation}
\begin{equation}
m_\phaseIV= - \frac{1}{\ttauL} \Big(\frac{\tilde{L}_*}{1-\tilde{N}_*}\Big);
\end{equation}
\end{subequations}
and similarly for the rest.
A convenient viewpoint on the condition is that
\begin{equation}
  m_\phaseph= \ttauL m_{\phaseph,0}
\end{equation}
where $m_{\phaseph,0}$	is the slope of the line from
$(\tilde{N}_*,\tilde{L}_*)$ to the appropriate corner
of the square.

Meanwhile, the slopes of the $\dot{\tilde{N}}=0$ and 
$\dot{\tilde{L}}=0$ isoclines are
\begin{subequations}
\begin{eqnarray}
m_N &\equiv& \tilde{C}_{NN}/\tilde{C}_{NL}, \\
m_L &\equiv& \tilde{C}_{LN}/\tilde{C}_{LL}.
\end{eqnarray}
\end{subequations}
Evidently, the additional (and sufficient) condition
to get a spiraling behavior is that $m_N < m_\phaseIII, m_\phaseI < m_L$.
Since $\ttauL<1$ and $m_\phaseph$ is of order $1/\ttauL$
for $\rm ph=I,II,III,IV$,  it follows that $m_L$ is typically
large and hence $C_{LL}$ must be relatively small.

Solving one linear equation for each phase of the cycle, we find
after four isocline intersections that
$(\delta \tilde{N}_{i+4},\delta \tilde{L}_{i+4})=\Lambda (\delta \tilde{N}_i,\delta \tilde{L}_i)$, where
\begin{equation}
  \Lambda \equiv \frac{R_\phaseI R_\phaseIII}{R_\phaseII R_\phaseIV}
\end{equation}
where
\begin{equation}
   R_\phaseph \equiv \frac{m_L-m_\phaseph}{m_N-m_\phaseph}.
\end{equation}
for $\rm ph= I,II,III, IV$.
Thus, the fixed point $(\tilde{N}_*,\tilde{L}_*)$ is stable if and only if 
$\Lambda< 1$.
An interesting special case is when 
$(\tilde{N}_*,\tilde{L}_*)=(1/2,1/2)$: 
in that case, $\Lambda\equiv 1$ so the fixed point is always marginal.

To get a better understanding of the stability, consider the
case that $m_N \ll m_\phaseI, ..., m_\phaseIV \ll m_L$.
Then 
\begin{equation}
   R_\phaseph \approx - \frac{m_L}{m_\phaseph} \Big(1-
\frac{m_\phaseph}{m_L} + \frac{m_N}{m_\phaseph}\Big).
\end{equation}
Noting that $m_\phaseI m_\phaseIII/m_\phaseII m_\phaseIV \equiv 1$,
we get
\begin{eqnarray}
   \Lambda &\approx& 1- \frac{1}{\ttauL m_L} 
\Big(m_{\phaseI,0} + 
    m_{\phaseIII,0} - m_{\phaseII,0} - m_{\phaseIV,0}\Big) +
    \nonumber\\
    &&+ \ttauL m_N 
\Big(m_{\phaseI,0}^{-1} + m_{\phaseIII,0}^{-1} 
-m_{\phaseII,0}^{-1} - m_{\phaseIV,0}^{-1}\Big).
\end{eqnarray}
The sums in parentheses do not have a factor of $\ttauL$;
they depend only on $(\tilde{N}_*,\tilde{L}_*)$.
Both sums tend to be positive 
when the
fixed point is farther from 
the line $\tilde{L}=\tilde{N}$
than it is from the line $\tilde{L}=1-\tilde{N}$, or both negative in
the opposite situation.
Thus, in the former situation,
the fixed point tends to be unstable when
    \begin{equation}
     (m_L m_N)^{1/2} < \ttauL.
    \end{equation}
In the opposite situation, the fixed point tends to be 
unstable when the inequality is in the other direction.

\LATER{I need to check on the plus or minus sign on $m_\phaseph$}

The oscillation period $T$ can be inferred from the orbit
on the $(N,L)$ plane by going back to equation \eqref{eq:N-t-constant-s}.
Note that $\dot{\tilde{N}}$ and $\dot{\tilde{L}}$ do not go to
zero approaching the isocline line, but rather they approach a
positive or negative constant coming from the respective sides.
In other words, the four legs of the cycle on the $(\tilde{N},\tilde{L})$ plane
are each traversed at a roughly constant speed.  Hence a 
cycle that forms a small loop around $(\tilde{N}_*,\tilde{L}_*)$,
as happens just beyond the instability, has its period $T$ proportional
to its oscillation amplitude.

It follows that if the parameters are close to the instability
threshold, we can construct a spatially periodic solution
of intervals containing oscillations which are shifted in
phase with respect to each other, provided that the spatial
period is large compared to the distance that either signal
can diffuse in time $T$: locally, this situation is equivalent
to the uniform one, since the signal from contrasting 
intervals does not have time to propagate.  It follows that
periodic traveling waves are also possible, as well as 
a standing wave made up of alternating domains that each 
oscillate similar to the uniform oscillations, but in 
opposite phase to each other.

\section{Conclusion}

In this work, we first 
sketched (Sec.~\ref{sec:2D-model})
the realistic two-dimensional model developed in \cite{xu},
which exhibits a uniformly moving posterior-to-anterior front of
Nodal production 
on one of the two stripes representing the lateral plate mesoderm,
followed by a front of Lefty production on the midline,
just like the experimental observation~\cite{wang-yost}, and which 
is known to propagate the left/right asymmetry from the tailbuds
through the entire embryo.  We then set up a version of this
model in one dimension (Sec.~\ref{sec:1D-model}), representing the anterior-to-posterior axis,
which captures most properties of front propagration.
The one-dimensional space represents a combination of the 
left LPM and the midline, and its main qualitative defect
compared to the two-dimensional system is that it misses the
lag time for the Nodal signal $N$ to diffuse from the LPM to
the midline, or for the Lefty signal $L$ to diffuse from the
midline to the LPM.  The key parameters of the model were
(1) coefficients [Sec.~\ref{sec:N-L-production}] quantifying 
how much Nodal promotes the two signals, and how much
Lefty inhibits them, and (2) different diffusion constants 
and/or degradation times for the two signals, allowing
us to define a typical length scale for either one, over which 
its concentration varies in front of and behind a sharp
spatial step of the production; all in all, we found seven
dimensionless ratios, each of which is a nontrivial parameter
(Sec.~\ref{sec:non-dim}).

We then set out to classify, analytically, the possible behaviors 
of the model in a steady state, and see how the depended on the
parameter values.  A key aid in identifying the regime is make
a two dimensional plot (Sec.~\ref{sec:N-L-plane}) of how the homogeneous 
concentrations $(N,L)$ would evolve: the very existence of a front depends on 
having a fixed point on this plane representing nonzero Nodal production,
alongside the fixed point of zero production (which is always present).
Our main focus was on uniformly moving
fronts, but in fact one can specialize (without real loss of
generality, and with considerable mathematical simplification) 
to the special case of zero front velocity [Sec.~\ref{sec:traveling-wavefront}].
The key special feature we found was the possibility that, instead
of showing a sharp front where the production rapidly switches from
zero to maximum, the Lefty production (on the midline in the real
embryo) might be varying continuously over an extended interval
(Sec.~\ref{sec:results}).
This was due to a sort of feedback or buffering, in which the Nodal and Lefty 
concentrations compensate exactly such that every point along this interval
is right at the threshold for Lefty production (``pinned'').

Finally, in Sec.~\ref{sec:oscillating}
we changed focus to the possibility of a temporally periodic behavior;
this parameter regime allows both the possibility of a steady limiting
state which is ``pinned'' (at threshold) for {\it both} Nodal and Lefty.
One may get oscillations if Lefty degrades much faster than Nodal,
and we characterized the exact mathematical criterion for the steady
state to be unstable to developing oscillations.

Having found the relationship of these behaviors to
the parameter values, in future work we will be able to
find the corresponding parameter regimes in the realistic
two-dimensional model, with the ultimate goal of finding a set that gives
the observed phenotypes.  We obtained many behaviors that
are {\it not} seen experimentally.
(It is still uncertain whether the Lefty production has a sharp front
or an extended one like the ``pinned'' case.)  This is a 
positive feature of the model, for it will allow the exclusion of
whole domains of parameter values.

Our results might motivate further studies of the possible 
behaviors of propagating L/R
asymmetry in the zebrafish and other embryos. 
The pertinent measurements in real systems that relate to
phenomena in this model are 
\begin{itemize}
\item[(1)]
The measured front velocity constrains the velocity scale
$v_N$.
\item[(2)]
The sharpness of the $N$ and $L$ concentration fronts gives the length scales 
$l_N$ and $l_L$ and/or indicates the possibility of  the
``pinned'' interval; however, experiments that measure mRNA
are in effect probing the production rate and do not
access $l_N$ and $l_L$ directly.
\item[(3)]
The lag between the Nodal front and the Lefty front,
gives information on the difference between Nodal and
Lefty threshold parameters.
Experimentally the Lefty front lags, but many parameter
sets give a Lefty front that spatially leads (even though
causally it is downstream from Nodal), so observation
can limit parameters to a certain window.
\end{itemize}
In addition, they suggest how changing a parameter
(due to a mutation or some external perturbation) could
give an exotic result, such as non-monotonic front.

What the present study did {\it not} address is how the
Nodal production is first initiated, and why it only
happens along the left LPM. That depends on the
{\it two-dimensional} geometry, e.g. the relative
separations of the tailbud from the foot of the 
LPM stripes and the midline stripe of Oep-producing cells.
Furthermore, it is possible the real
system could propagate, e.g., a finite pulse of Nodal
that terminates, but that will be seen only if such a
signal is produced at the posterior end.  
These questions must be left to a proper two-dimensional
study.

%


\section*{SUPPLEMENTARY MATERIAL}


This work was supported by the Department of Energy grant DE-FG02-89ER-45405 (C.L.H.)
and  National Institute of Child Health \& Human Development grant 
R01-HD048584 (B. X. and R.D.B.).

\end{document}